\documentclass[procedia]{easychair}

\usepackage{doc}
\usepackage{makeidx}
\usepackage{bm}
\usepackage{color}
\usepackage{braket}

\usepackage{ulem}

\def\procediaConference{99th Conference on Topics of
  Superb Significance (COOL 2014)}

\title{Analysis of Charge-spin-orbital Fluctuations by {\it Ab Initio} Calculation and Random Phase Approximation: Application to Non-coplanar Antiferromagnet Cd$_2$Os$_2$O$_7$}

\titlerunning{Analysis of Charge-spin-orbital Fluctuations by \ldots}

\author{
  Amane Uehara\inst{1}
\and
  Hiroshi Shinaoka\inst{2,3}
\and
  Yukitoshi Motome\inst{1}
}

\institute{
  Department of Applied Physics, University of Tokyo, Tokyo 113-8656, Japan
\and
  Theoretische Physik, ETH Z\"{u}rich, 8093 Z\"{u}rich, Switzerland
\and
  Department of Physics, University of Fribourg, 1700 Fribourg, Switzerland
}

\authorrunning{Uehara, Shinaoka and Motome}

\begin{document}

\maketitle

\keywords{{\it ab initio} calculation, random phase approximation, strong electron correlation, Cd$_2$Os$_2$O$_7$, spin fluctuation, all-in all-out}

\begin{abstract}
We present a systematic analysis on the basis of {\it ab initio} calculations and many-body perturbation theory for clarifying the dominant fluctuation in complex charge-spin-orbital coupled systems.
For a tight-binding multiband model obtained from the maximally-localized Wannier function analysis of the band structure by the local density approximation, we take into account electron correlations at the level of random phase approximation.
To identify the dominant fluctuation, we carry out the eigenmode analysis of the generalized susceptibility that includes all the multiple degrees of freedom: charge, spin, and orbital.
We apply this method to the paramagnetic metallic phase of a pyrochlore oxide Cd$_2$Os$_2$O$_7$, which shows a metal-insulator transition accompanied by a peculiar noncoplanar antiferromagnetic order of all-in all-out type.
We find that the corresponding spin fluctuation is dominantly enhanced by the on-site Coulomb repulsions in the presence of strong spin-orbit coupling and trigonal crystal field splitting.
Our results indicate that the combined method offers an effective tool for the systematic analysis of potential instabilities in strongly correlated electron materials.
\end{abstract}

\section{Introduction}

The interplay of charge, spin, and orbital degrees of freedom under the electron correlation has been one of the central issues in condensed matter physics~\cite{Imada-Fujimori-Tokura-1998-MIT, Tokura-Nagaosa-orbital-2000, E.Dagotto-SCES-2005}.
It brings about fascinating phenomena in a wide range of physical properties, including magnetism, optics, and transport.
Among them, the properties induced by symmetry breaking and associated fluctuations have been of primary important, since they lead to a dramatic change of the equilibrium states and a gigantic response to an external field.
Despite a long history of research, a systematic theoretical study of complicated entanglement between the multiple degrees of freedom remains as a challenging issue in condensed matter physics. 

For clarifying the underlying mechanism in these complicated phenomena, it is helpful to identify which degree of freedom of electrons plays a role.
An efficient way is to analyze the fluctuations.
For a continuous phase transition, when approaching the critical point from the high-symmetry phase, a particular fluctuation diverges corresponding to the symmetry breaking, which provides the information of the relevant degree of freedom in the phase transition.
Even in the case of a first-order phase transition, the analysis of enhanced fluctuations might give useful information.
Thus, identifying the dominant fluctuation in the high-symmetry phase is significantly important for the understanding of phase transition phenomena.

In this paper, we present a systematic study of the dominant fluctuation in the charge, spin, and orbital coupled systems.
Our method is twofold: the construction of the tight-binding model by {\it ab initio} calculations and the perturbation theory for electron correlations.
In the first step, we perform band structure calculations on the basis of the relativistic local density approximation (LDA), and estimate the tight-binding parameters by the maximally-localized Wannier function (MLWF) analysis.
In the second step, we take into account the effect of electron correlations at the level of random phase approximation (RPA), and calculate the generalized susceptibility that includes all charge, spin, and orbital degrees of freedom.
To identify the dominant fluctuation, we carry out the eigenmode analysis of the generalized susceptibility while changing the model parameters as well as temperature.

\begin{figure}[htbp]
  \centering
  \includegraphics[width=.7\columnwidth]{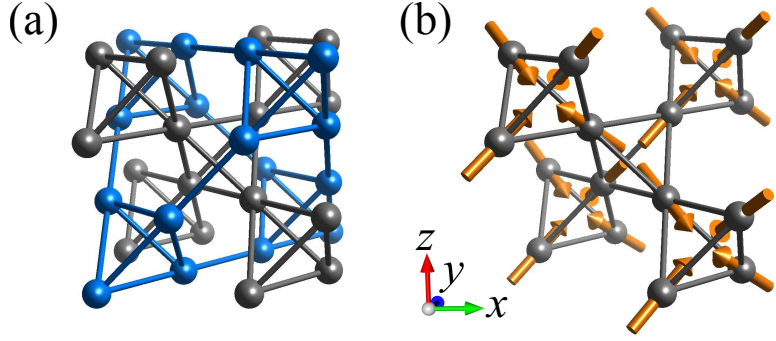}
  \caption{\label{fig:pyrochlore}
    (color online).
    (a) Schematic picture of the lattice structure of Cd$_2$Os$_2$O$_7$.
    The blue and gray spheres denote Cd and Os, respectively.
    (b) The pyrochlore lattice structure composed of Os cations.
    The allows represent the magnetic pattern of the all-in all-out order below $T_{\rm c}$.
    The spins point in the local [111] directions, which connect the centers of neighboring Os$_4$ tetrahedra.
    All-in and all-out tetrahedra appear alternatively, resulting in the $\bm{q}=0$ noncoplanar magnetic order.
  }
\end{figure}

We apply the technique to a pyrochlore oxide Cd$_2$Os$_2$O$_7$ and discuss the results in comparison with those in the previous experimental and theoretical studies.
Here, we briefly summarize the physical properties of this compound.
The lattice structure of Cd$_2$Os$_2$O$_7$ is schematically shown in Fig.~\ref{fig:pyrochlore}(a).
Cd$^{2+}$ cations are nonmagnetic, while Os$^{5+}$ cations have three electrons in $t_{\rm 2g}$ orbitals per site, and comprise the pyrochlore lattice, as shown in Fig.~\ref{fig:pyrochlore}(b).
This material exhibits a metal-insulator transition at $T_{\rm c}=225$~K~\cite{A.W.Sleight-1974-Cd2Os2O7, Mandrus-2001-Cd2Os2O7}.
The phase transition is of second order, and no structural changes are observed.
The transition is accompanied by a noncoplanar antiferromagnetic order of the so-called all-in all-out type, as shown in Fig.~\ref{fig:pyrochlore}(b)~\cite{Yamaura-2012-Cd2Os2O7}.
This magnetic order occurs with the wave number $\bm{q}=0$, which does not break the symmetry of the lattice. 
On the theoretical side, the electronic structure of this material was discussed by {\it ab initio} calculations including relativistic effects~\cite{H.Harima-Cd2Os2O7-2002, D.J.Singh-Cd2Os2O7-2002}.
The peculiar all-in all-out type magnetic order as well as the metal-insulator transition was explained by including the electron correlations by the relativistic LDA+$U$ method~\cite{Shinaoka-2012-Cd2Os2O7}.

\section{Model and Method \label{sec:model-method}}

\subsection{Model construction}

We start with {\it ab initio} calculations based on the density functional theory with LDA for the target material~\cite{P.Hohenberg-1964-LDA, W.Kohn-1965-LDA}: Cd$_2$Os$_2$O$_7$ in the current case.
In the LDA calculation, we use a fully-relativistic {\it ab initio} computational code, QMAS (Quantum Materials Simulator)~\cite{QMAS}.
The relativistic effects are fully taken into account.
We employ the projector augmented wave method~\cite{PAW}.
The computational condition is the same as that in the previous study for Cd$_2$Os$_2$O$_7$~\cite{Shinaoka-2012-Cd2Os2O7}.

Based on the LDA band structure, we construct the multiband tight-binding model for the relevant orbitals: in the current case, the $t_{\rm 2g}$ orbitals of $5d$ electrons in Os cations (see Fig.~\ref{fig:band}).
The multiband Hamiltonian is given by
$\mathcal{H} = \mathcal{H}_0 + \mathcal{H}_{\rm int}$, where $\mathcal{H}_0$ and $\mathcal{H}_{\rm int}$ denote the one-body and two-body part, respectively.
In order to estimate the tight-binding parameters in the one-body part $\mathcal{H}_0$, we adopt the MLWF analysis of the LDA results~\cite{Marzari-1997-MLWF,Souza-2001-MLWF}, as follows.
Here, $\mathcal{H}_0$ consists of three parts, namely, electron transfer $\mathcal{H}_{\rm t}$, crystal-field splitting $\mathcal{H}_{\rm CF}$, and spin-orbit interaction $\mathcal{H}_{\rm SOI}$.
$\mathcal{H}_{\rm t}$ is given by
\begin{align}
\mathcal{H}_{\rm t}
= \sum_{\bm{k} \in {\rm 1BZ}} \sum_{\zeta \bm{\rho} \zeta' \bm{\rho}'}
t_{\zeta \bm{\rho}, \zeta' \bm{\rho}'} ( \bm{k})
\sum_{\sigma}
c_{\zeta  \sigma \bm{\rho}  \bm{k}}^{\dagger}
c_{\zeta' \sigma \bm{\rho}' \bm{k}}
\label{eq:Ht},
\end{align}
where $c_{\zeta \sigma \bm{\rho} \bm{k}}$ and $c_{\zeta \sigma \bm{\rho} \bm{k}}^{\dagger}$ denote the Fourier transform of the annihilation and creation operators, respectively, for the obtained MLWF $\psi_{ \zeta \sigma \bm{R}+\bm{\rho} }( \bm{r} )$; $\zeta$, $\sigma$, $\bm{R}$, and $\bm{\rho}$ denote the orbital, spin, unit cell, and sublattice in the unit cell, respectively.
The summation of the wave number $\bm{k}$ is taken in the first Brillouin zone for the Bravais lattice defined by $\{\bm{R}\}$.
The hopping matrix element $t_{\zeta \bm{\rho}, \zeta' \bm{\rho}'} ( \bm{k})$ is the Fourier transform of the overlap integral given by
\begin{align}
t_{\zeta,\zeta'} \left[ ( \bm{R} + \bm{\rho}) - ( \bm{R}' + \bm{\rho}')\right] =
\int d\bm{r}
\psi_{ \zeta \sigma \bm{R}+\bm{\rho} }^{*}( \bm{r} )
\left( - \frac{\hbar^2}{2m} \nabla^2 \right)
\psi_{ \zeta' \sigma' \bm{R}'+\bm{\rho}' }( \bm{r} )
\label{eq:def-transfer-matrix}.
\end{align}
For $\mathcal{H}_{\rm CF}$, we include the dominant trigonal distortion of each OsO$_6$ octahedra~\cite{Mandrus-2001-Cd2Os2O7}, as
\begin{align}
\mathcal{H}_{\rm CF}
&= \Delta \times
\begin{pmatrix}
0 & 1 & 1 \cr
1 & 0 & 1 \cr
1 & 1 & 0
\end{pmatrix}
\label{eq:HCF},
\end{align}
where the basis is taken as ($d_{yz}, d_{zx}, d_{xy}$).
Meanwhile, $\mathcal{H}_{\rm SOI}$ is given by
\begin{align}
\mathcal{H}_{\rm SOI} &= \lambda \bm{l} \cdot \bm{s} =
\frac{\lambda}{2} \times
\begin{pmatrix}
 0         & -i\sigma_z &  i\sigma_y \cr
 i\sigma_z &  0         & -i\sigma_x \cr
-i\sigma_y &  i\sigma_x &  0
\end{pmatrix}
\label{eq:HSOI},
\end{align}
where the basis is taken as in Eq.~\eqref{eq:HCF} including the spin degree of freedom explicitly, and $\sigma_{x}$, $\sigma_{y}$, and $\sigma_{z}$ denote the Pauli matrices for spin indices.
We estimate the coupling constants $\Delta$ and $\lambda$ by using the MLWFs as
\begin{align}
\Delta &= \langle d_{xy} \uparrow | \mathcal{H}^{\rm LDA} | d_{yz} \uparrow \rangle
\label{eq:HCF-estimate},
\\
\frac{\lambda}{2} &= \langle d_{xy} \uparrow | \mathcal{H}^{\rm LDA} \Ket{d_{yz} \downarrow}
\label{eq:SOI-estimate},
\end{align}
where $\mathcal{H}^{\rm LDA}$ is the Hamiltonian in the relativistic Kohn-Sham equation.

For the two-body part of the Hamiltonian, we consider the on-site electron correlation defined by
\begin{align}
\mathcal{H}_{\rm int} = \sum_{\zeta\mu\zeta'\mu'} U_{\zeta\mu,\zeta'\mu'} \sum_{\sigma_1 \sigma_2} \sum_{\bm{r}}
c_{\zeta \sigma_1 \bm{r}}^{\dagger}
c_{\mu   \sigma_2 \bm{r}}^{\dagger}
c_{\mu'  \sigma_2 \bm{r}}
c_{\zeta'\sigma_1 \bm{r}}
\label{eq:Hint}.
\end{align}
For simplicity, we assume that the MLWFs possess the same symmetry as the atomic orbitals.
Then, the coupling constant is simplified into the form of
\begin{align}
U_{\zeta\mu,\zeta'\mu'} = (U-2J_{\rm H}) \delta_{\zeta \zeta'} \delta_{\mu \mu'}
+ J_{\rm H} (\delta_{\zeta \mu} \delta_{\zeta' \mu'} + \delta_{\zeta \mu'} \delta_{\mu \zeta'} )
\label{eq:simplified-U},
\end{align}
where $U$ and $J_{\rm H}$ denote the intra-orbital Coulomb interaction and the Hund's-rule coupling, respectively.
Here, we assume the rotational symmetry of the Coulomb interaction; $U=U'+2J_{\rm H}$ and $J_{\rm H}=J_{\rm pair}$, where $U'$ and $J_{\rm pair}$ are the inter-orbital Coulomb interaction and the pair hopping between different orbitals, respectively.

\subsection{Random phase approximation and eigenmode analysis}
\label{subsec:rpa}

We study charge, spin, and orbital fluctuations in the obtained multiband model by calculating the generalized susceptibility.
First, we calculate the bare susceptibility $\chi^{(0)}$ from the one-body part of the Hamiltonian $\mathcal{H}_0$, which is given by
\begin{align}
\chi^{(0)}_{\alpha\beta\bm{\rho},\alpha'\beta'\bm{\rho}'}( q ) =
- T \sum_{\bm{k} \in {\rm 1BZ}} \sum_{\omega_k}
\mathcal{G}^{(0)}_{\alpha'\bm{\rho}',\alpha \bm{\rho} }( k     )
\mathcal{G}^{(0)}_{\beta  \bm{\rho} ,\beta' \bm{\rho}'}( k + q )
\label{eq:chi0},
\end{align}
where $\alpha$ denotes the orbital and spin indices as $\alpha = (\zeta_{\alpha}, \sigma_{\alpha})$, and $T$ denotes the temperature.
Here, $\mathcal{G}^{(0)}_{\alpha'\bm{\rho}',\alpha \bm{\rho} }(q)$ is the noninteracting Green function, where $q = (\bm{q}, \omega_{q})$; $\omega_{q}$ is the Matsubara frequencies.

Next, we take into account the effect of two-body part $\mathcal{H}_{\rm int}$ at the level of RPA.
For this purpose, we define the vertex function as
\begin{align}
\mathcal{U}_{ \alpha  \beta  \bm{\rho} , \beta' \alpha' \bm{\rho}'} = (
  U_{\zeta_{\alpha } \zeta_{\beta },\zeta_{\beta' } \zeta_{\alpha'}} \delta_{\sigma_{\alpha } \sigma_{\beta' }} \delta_{\sigma_{\beta } \sigma_{\alpha'}} +
  U_{\zeta_{\alpha } \zeta_{\beta'},\zeta_{\alpha'} \zeta_{\beta  }} \delta_{\sigma_{\alpha } \sigma_{\alpha'}} \delta_{\sigma_{\beta'} \sigma_{\beta  }}
) \delta_{\bm{\rho}  \bm{\rho}'}
\label{eq:vertex},
\end{align}
 and calculate the generalized susceptibility in the matrix form of
\begin{align}
\chi^{\rm RPA}(q) =
\left[I - \chi^{(0)}( q) \mathcal{U} \right]^{-1} \chi^{(0)}( q)
\label{eq:Dyson},
\end{align}
where $I$ denotes the unit matrix.

We can find the dominant fluctuation by studying the eigenvalues and eigenvectors of the matrix $\chi^{\rm RPA}(q)$, as follows.
According to the fluctuation-dissipation theorem, the generalized susceptibility satisfies the relation:
\begin{align}
\delta \left\langle c_{\alpha\bm{\rho}\bm{q}}^{\dagger} c_{\beta\bm{\rho}\bm{q}} \right\rangle=
\sum_{\alpha'\beta'\bm{\rho}'} \chi_{\alpha\beta\bm{\rho} ,\alpha'\beta'\bm{\rho}'}^{\rm RPA}(q) h_{\alpha'\beta'\bm{\rho}'}(q),
\label{eq:fluctuation-dissipation}
\end{align}
where $\delta \left\langle A \right\rangle$ represents the deviation of the thermal average of $A$ induced by a generalized external field $h_{\alpha'\beta'\bm{\rho}'}(q)$.
If the external field is parallel to the $\kappa$th eigenvector of $\chi^{\rm RPA}$, $e_{\alpha\beta\bm{\rho}}^{\kappa}$, Eq.~\eqref{eq:fluctuation-dissipation} reads
\begin{align}
\delta^{\kappa} \left\langle c_{\alpha\bm{\rho}\bm{q}}^{\dagger} c_{\beta\bm{\rho}\bm{q}} \right\rangle \propto x^{\kappa}(q) e_{\alpha\beta\bm{\rho}}^{\kappa}(q),
\label{eq:fluctuation-dissipation-parallel}
\end{align}
where $\delta^{\kappa} \left\langle A \right\rangle$ represents $\delta \left\langle A \right\rangle$ in the $\kappa$th eigenmode, and $x^{\kappa}(q)$ is the $\kappa$th eigenvalue.
This indicates that $\delta^{\kappa} \left\langle c_{\alpha\bm{\rho}\bm{q}}^{\dagger} c_{\beta\bm{\rho}\bm{q}} \right\rangle$ also becomes parallel to $e_{\alpha\beta\bm{\rho}}^{\kappa}$.
Thus, the eigenvector and eigenvalue correspond to the normal mode and amplitude of the fluctuation, respectively.
Therefore, the analysis of the eigenmodes provides the information of charge, spin, and orbital components of the fluctuations.
For example, the spin fluctuation $\delta \bm{s}_{\bm{\rho}}^{\kappa}(q)$ in the $\kappa$th eigenmode at the sublattice $\bm{\rho}$ is written as
\begin{align}
\delta \bm{s}_{\bm{\rho}}^{\kappa}(q)
=\delta^{\kappa} \left\langle \sum_{\zeta \sigma\sigma'} c_{\zeta \sigma \bm{\rho}\bm{q}}^{\dagger} \bm{\sigma}_{\sigma\sigma'}c_{\zeta \sigma' \bm{\rho}\bm{q}} \right\rangle
\propto\sum_{\zeta \sigma\sigma'} \bm{\sigma}_{\sigma\sigma'} e_{(\zeta, \sigma) (\zeta \sigma') \bm{\rho}}^{\kappa}(q),
\label{eq:fluctuation-delta-spin}
\end{align}
where $\bm{\sigma}$ denotes the vector of Pauli matrices; $\bm{\sigma} = (\sigma_x, \sigma_y, \sigma_z)$.

\section{Application to Cd$_2$Os$_2$O$_7$}

\begin{figure}[htbp]
  \centering
  \includegraphics[width=1.\columnwidth]{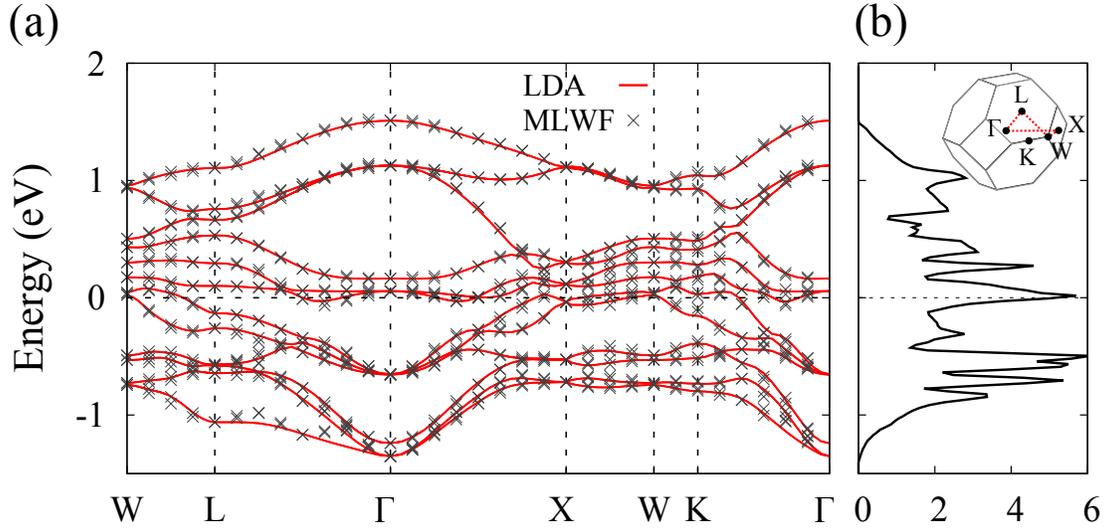}
  \caption{\label{fig:band}
    (color online).
    (a) Electronic band structure for Cd$_2$Os$_2$O$_7$ near the Fermi level.
    The curves denote the results obtained by the relativistic LDA calculations, while the crosses show those by the tight-binding Hamiltonian $\mathcal{H}_0$ obtained by the MLWF  analysis.
    (b) The density of states calculated  for $\mathcal{H}_0$.
    In (a) and (b), the Fermi level is set to zero.
    The inset in (b) represents the first Brillouin zone for the pyrochlore lattice structure.
  }
\end{figure}

We apply the method in the previous section to a pyrochlore oxide Cd$_2$Os$_2$O$_7$.
The band structure for this compound was calculated in the previous study~\cite{Shinaoka-2012-Cd2Os2O7} for the experimental lattice structure at 180~K~\cite{Mandrus-2001-Cd2Os2O7}.
Note that the system does not show a structural change at the metal-insulator transition at $225$~K.
Figure~\ref{fig:band}(a) shows the relativistic LDA band structure.
The band structure near the Fermi level consists mainly of $t_{\rm 2g}$ orbitals of the Os cations, which are well separated from other bands.
We carry out the MLWF analysis for the $t_{\rm 2g}$ bands and estimate the tight-binding parameters in $\mathcal{H}_0$.
As shown in Fig.~\ref{fig:band}(a), the tight-binding band structure calculated by $\mathcal{H}_0$ well reproduces the LDA result.
In this $5d$ electron system, the spin-orbit interaction is relevant: $\lambda = 332$~meV, which is larger than the largest matrix element in the nearest neighbor hopping, $|t|=178$~meV.
The trigonal crystal field splitting is also substantial: $\Delta = 96.6$~meV. 
Figure~\ref{fig:band}(b) shows the density of states obtained by the MLWFs.
There is a sharp peak of DOS at the Fermi level.
In the previous relativistic LDA + $U$ study, it was discussed that the Coulomb repulsion $U$ suppresses the DOS and induces the all-in all-out type magnetic order~\cite{Shinaoka-2012-Cd2Os2O7}.
In the following, we examine the electron fluctuation on the basis of the LDA band structure by using the eigenmode analysis of the generalized susceptibility.

\begin{figure}[htbp]
  \centering
  \includegraphics[width=.7\columnwidth]{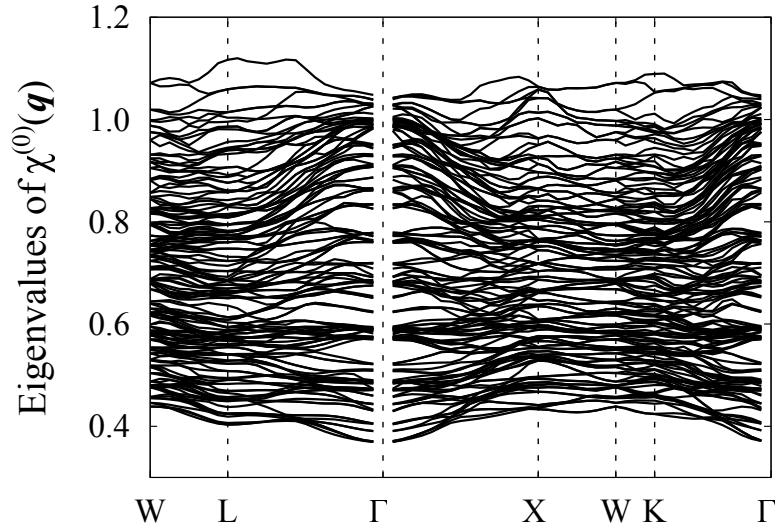}
  \caption{\label{fig:chi0}
    Eigenvalues of the bare susceptibility $\chi^{(0)}(\bm{q})$ calculated by using the tight-binding parameters obtained from the MLWFs.
  }
\end{figure}

First, we compute the bare susceptibility $\chi^{(0)}$ in Eq.~\eqref{eq:chi0}.
Figure~\ref{fig:chi0} shows all the eigenvalues of the static component, $\chi^{(0)}(\bm{q}) = \chi^{(0)}(\bm{q}, \omega=0)$.
In the calculations, we take $32^3$ $\bm{q}$ points in the first Brillouin zone and set the temperature to 10~K.
Although the temperature is much lower than $T_{\rm c} = 225$~K, we confirmed that the following results are qualitatively similar when the system is in the paramagnetic state in the vicinity of instability toward the magnetic ordering discussed below.
The number of the eigenmodes is 144: we show the sublattice diagonal components of $\chi^{(0)}(\bm{q})$ with three orbital and two spin components, namely, $144 = 4 \times (3 \times 2)^2$.
Note that the eigenvalues at $\bm{q}=0$ are not available precisely in the present scheme due to the technical reason.
As shown in Fig.~\ref{fig:chi0}, the largest eigenmode has a broad peak around the L point, while it is rather suppressed around the $\Gamma$ point.
However, as described below, this wave number dependence is changed considerably by electron correlations.

\begin{figure}[htbp]
  \centering
  \includegraphics[width=1.\columnwidth]{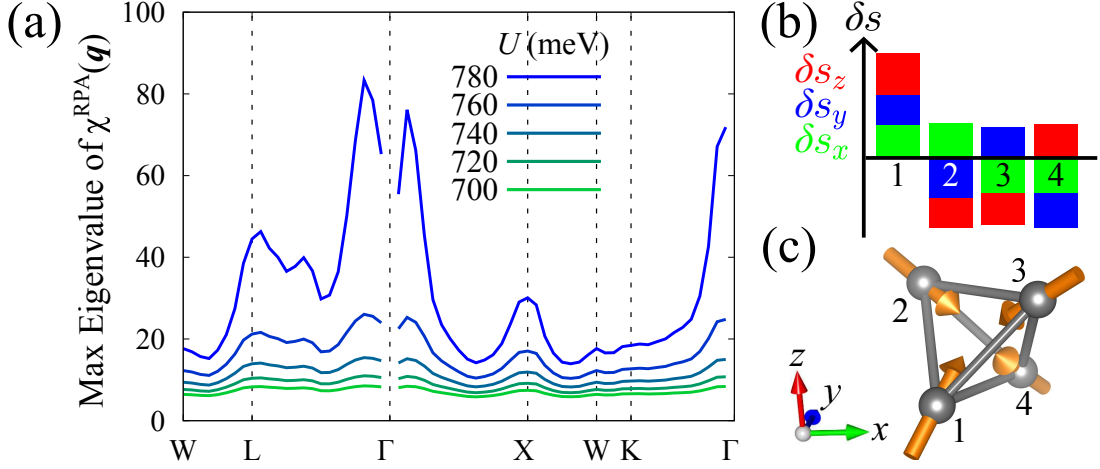}
  \caption{\label{fig:rpa}
    (color online).
    (a) $U$ dependence of the maximum eigenvalues of $\chi^{\rm RPA}(\bm{q})$.
    We change the intra-orbital Coulomb interaction $U$, while keeping $J_{\rm H}/U=0.2$ and $T=10$~K.
    (b) Spin fluctuations corresponding to the maximum eigenmode near the $\Gamma$ point, at $\bm{q} = \bm{q}_0 = (\pi/16, \pi/16, \pi/16)$.
    The calculations are done for $U=780$~meV and $J_{\rm H}=156$~meV.
    The vertical axis represents the amplitudes of spin fluctuations in an arbitrary unit.
    Green, blue, and red bars represent the amplitudes of the $x$, $y$, and $z$ spin components, respectively.
    The numbers 1-4 in the horizontal axis denote the sublattices.
    (c) Schematic visualization of the spin fluctuations in the maximum eigenmode in (b).
  }
\end{figure}

Next, we compute the generalized susceptibility at the level of RPA by Eq.~\eqref{eq:Dyson}.
Figure~\ref{fig:rpa}(a) shows the maximum eigenvalues of $\chi^{\rm RPA}(\bm{q})$.
We show the data while changing the intra-orbital Coulomb interaction $U$ from 700~meV to 780~meV with keeping its ratio to the Hund's-rule coupling as $J_{\rm H}/U=0.2$.
Other parameters are taken to be the same as in Fig.~\ref{fig:chi0}.
As shown in Fig.~\ref{fig:rpa}(a), the eigenvalues around the $\Gamma$ point are strongly enhanced as $U$ increases.

We investigate the nature of dominant fluctuation by using the eigenmode analysis introduced in Sec.~\ref{subsec:rpa}.
Here, we consider the eigenmode at $\bm{q}_0 = (\pi/16, \pi/16, \pi/16)$, which is the smallest wave number along the $\Gamma$-L line.
We find that, in the dominant eigenmode, spin fluctuations defined by Eq.~\eqref{eq:fluctuation-delta-spin} have much larger amplitudes compared to charge and orbital components.
Figure~\ref{fig:rpa}(b) shows the $x$, $y$, and $z$ components of the spin fluctuations at $\bm{q}_0$.
We find that all the three components have almost similar amplitudes, indicating that
the directions of the spin fluctuations are roughly in the $(1,1,1)$, $(1,-1,-1)$, $(-1,1,-1)$, and $(-1,-1,1)$ directions at the sublattices 1, 2, 3, and 4, respectively.
The spin fluctuation is of all-in all-out type, whose schematic visualization is shown in Fig.~\ref{fig:rpa}(c).
Thus, this peculiar spin fluctuation, which breaks the rotational symmetry in spin space, is caused by the electron correlation in the presence of the strong spin-orbit coupling and the trigonal crystalline electric field.
The importance of the spin-orbit coupling and trigonal crystalline electric field was also pointed out in the previous LDA+$U$ theory~\cite{Shinaoka-2012-Cd2Os2O7}.

\section{Discussions and Concluding Remarks}

Let us discuss our results in comparison with the experiments and the previous theory for Cd$_2$Os$_2$O$_7$.
In our results, the fluctuation is dominantly enhanced around the $\Gamma$ point, as shown in Fig.~\ref{fig:rpa}(a).
In addition, the dominant fluctuation is in the spin channel and of all-in all-out type, as shown in Figs.~\ref{fig:rpa}(b) and \ref{fig:rpa}(c).
The results are consistent with the experiments in Cd$_2$Os$_2$O$_7$: a $\bm{q}=0$ magnetic order with all-in all-out type was observed associated with the metal-insulator transition~\cite{Yamaura-2012-Cd2Os2O7}.
We note that the peculiar magnetic order was reproduced by the relativistic LDA+$U$ method, in which the electron correlation is treated at the mean-field level and fluctuations are neglected~\cite{Shinaoka-2012-Cd2Os2O7}.
In this sense, our RPA study on fluctuations is complementary to the previous theory.

Our method is not specific to this compound: it provides a versatile tool to study dominant fluctuations in a wide range of materials. 
We can also argue, for clarifying the underlying physics, how the fluctuations are affected by controlling the model parameters, not only the electron correlations but also the one-body tight-binding parameters.
Moreover, the method can be more sophisticated by adopting other techniques to treat the electron correlations, such as the self-consistent fluctuation theory. 
These extensions of the model and method are left for future study.

We acknowledge fruitful discussions with T.~Misawa, M.~Udagawa, and Y.~Yamaji.
A.U. is supported by Japan Society for the Promotion of Science through Program for Leading Graduate Schools (MERIT).
The crystal structures are visualized by using VESTA 3~\cite{VESTA}.


\end{document}